\documentclass[11pt,letterpaper]{article}
\usepackage{mathrsfs}           % \mathscr
\usepackage{vmargin,fancyhdr}   % margins
\usepackage{algorithm}
\usepackage[noend]{algorithmic}

\usepackage{enumerate}

\usepackage{amsmath,amssymb}    % AMS
\usepackage{verbatim}           % comment environment
\usepackage{xspace}             % optional space
\usepackage{graphicx,float}     % figures
\usepackage{ifthen,calc}        % macros
\usepackage{textcomp}           % \textcent
\usepackage{fancybox}           % \Ovalbox
\usepackage{hhline}             % \hhline
\usepackage{float}              % \float

\usepackage[rflt]{floatflt}     % default is vflt
\usepackage{setspace}
\usepackage{subfigure}

\usepackage{color}
\definecolor{ForestGreen}{rgb}{0.1333,0.5451,0.1333}
\usepackage[letterpaper,%dvips,
            colorlinks,linkcolor=ForestGreen,citecolor=ForestGreen,
            backref,
            bookmarks,bookmarksopen,bookmarksnumbered]
           {hyperref}
\setpapersize{USletter}
\setmarginsrb{1in}{.5in}        % left, top
             {1in}{0.5in}        % right, bottom
             {0in}{.5in}     % headheight, headsep
             {0in}{.5in}      % footheight, footskip
\newcommand{\showccc}[0]{0}
\newcommand{\ccc}[2][nothing]{% comment
  \ifthenelse{\showccc=0}{}{
    \ensuremath{^{\Lsh\Rsh}}\marginpar{\raggedright\tiny\textsf{%
        \ifthenelse{\equal{#1}{nothing}}{}{\textbf{#1}\\}#2}}}}

\newtheorem{theorem}{Theorem}[section]
\newtheorem{claim}[theorem]{Claim}

\newtheorem{corollary}[theorem]{Corollary}
\newtheorem{definition}[theorem]{Definition}

\newtheorem{lemma}[theorem]{Lemma}

\newcommand{\Proof}[0]{\smallskip\noindent\textit{\textbf{Proof}}\quad}
\newcommand{\Proofof}[1]{\smallskip\noindent\textit{\textbf{Proof of #1:}}\quad}

\newcommand{\QED}[0]{\hfill\ensuremath{\blacksquare}\medspace}

\newcommand{\dist}{\text{dist}}

\newcommand{\prob}[2][\mbox{}]{\ensuremath{\mathop{\text{\normalfont {\textbf{Pr}}}}_{#1}\left[#2\right]}}
\newcommand{\expct}[2][\mbox{}]{\ensuremath{\mathop{\text{\normalfont \textbf{E}}}_{#1}}\left[#2\right]}

\floatstyle{ruled}
\newfloat{algo}{tbp}{lop}%[section]
\floatname{algo}{Algorithm}

\usepackage{float}
\floatstyle{plain}\newfloat{myfig}{t}{figs}[section]
\floatname{myfig}{\textsc{Figure}}
\floatstyle{plain}\newfloat{myalg}{H}{algs}[section]
\floatname{myalg}{}
\DeclareMathOperator*{\poly}{poly}
\newcommand{\abs}[1]{\lvert#1\rvert}

\newcommand{\ceil}[1]{\lceil#1\rceil}
\newcommand{\Ceil}[1]{\left\lceil#1\right\rceil}
\newcommand{\reals}{\mathbb{R}}

\DeclareMathOperator*{\argmin}{arg\,min}

\title{
Improved Parallel Algorithms for Spanners and Hopsets
}

\author{
  Gary L.\ Miller\\
  Carnegie Mellon University\\
  \texttt{glmiller@cs.cmu.edu}\\
  \and
  Richard Peng\\
 MIT\\
 \texttt{rpeng@mit.edu}\\
  \and
  Adrian Vladu\\
  MIT\\
  \texttt{avladu@mit.edu}\\
  \and
  Shen Chen Xu\\
  Carnegie Mellon University\\
  \texttt{shenchex@cs.cmu.edu}
}

\begin{document}

\maketitle

\begin{abstract}

We use exponential start time clustering to design faster and more
work-efficient parallel graph algorithms involving distances.
Previous algorithms usually rely on graph decomposition routines with
strict restrictions on the diameters of the decomposed pieces.
We weaken these bounds in favor of stronger local probabilistic guarantees.
This allows more direct analyses of the overall process, giving:
\begin{itemize}
\item Linear work parallel algorithms that construct
spanners with $O(k)$ stretch and size $O(n^{1+1/k})$
in unweighted graphs, and size $O(n^{1+1/k} \log k)$ in weighted graphs.
\item Hopsets that lead to  the first parallel algorithm for approximating shortest
paths in undirected graphs with $O(m \poly\log n)$ work.
\end{itemize}

\end{abstract}

\thispagestyle{empty}

% \vfill \pagebreak

\section{Introduction}
\label{sec:intro}

Graph decompositions are widely used algorithmic routines.
They partition the graph to enable divide-and-conquer algorithms.
One form that has proven to be particularly useful is the low diameter decomposition:
the decomposed pieces should have small diameter, while few edges have
endpoints in different pieces.
Variants of the low diameter decomposition are used in algorithms for
spanners~\cite{Cohen98}, distance oracles~\cite{ThorupZ05}, and low stretch
embeddings~\cite{AlonKPW95,Bartal96,CohenMPPX14:arxiv}.

Early applications of the low diameter decomposition include distributed
algorithms by Awerbuch~\cite{Awerbuch85},
and low-stretch spanning trees by Alon et al.~\cite{AlonKPW95}.
Further study of low stretch tree embeddings led to a
probabilistic decomposition routine by Bartal~\cite{Bartal96}.
On an unweighted graph, this decomposition partitions the graph
so that only a $\beta$ fraction of the edges are cut, and the resulting
pieces have diameter $O(\beta^{-1} \log{n})$.

The development of parallel algorithms for finding tree embeddings~\cite{BlellochGKMPT14}
led to a parallel low diameter clustering routine using exponential start
times~\cite{MillerPX13}.
This routine then led to algorithms that generate tree embeddings
suitable for a variety of applications~\cite{CohenMPPX14:arxiv}.
The clustering algorithm itself has properties suitable for
% other applications such as
reducing the communication required in parallel connectivity
algorithms~\cite{ShunDB14}.
This suggests that exponential start time clusterings have a variety of other
applications in graph algorithms.

Graph decomposition routines are often invoked hierarchically,
leading to many levels, each refining the output of the previous one.
The $O(\log{n})$ discrepancy between the probability of edges being cut and
diameters of pieces in standard low diameter decomposition could then
accumulate through the levels.
To address this issue, recent algorithms using low diameter decompositions
usually require stronger properties at intermediate steps.

In this paper, we give an alternate approach based on the probabilistic
guarantees of the exponential start time clustering.
We regard the multiple levels as independent events, and analyze the output
probabilistically in each locality.
This weakens the interactions between the levels,
while still allowing us to analyze the final outcome.
We apply this method to several classical graph problems involving distances.

Spanners are sparse subgraphs that approximate distances in the original graph.
We show that one round of exponential start time clustering augmented with a
few edges leads to spanners.
This algorithm extends to the weighted setting by bucketing the edges
by weights, and then clustering them hierarchically.
This leads to an overhead of $O(\log k)$ in the size compared to the
optimal construction where $k$ is the stretch factor, and $O(\log{U})$ in
depth, where $U$ is the ratio between the maximum and minimum edge weights.

\begin{theorem}
\label{thm:mainSpanner}
\begin{sloppypar}
There exists an algorithm that given as input a graph $G$ and parameter $k\ge1$, finds with high
probability a subgraph $H$ in which shortest path distances are preserved up to
a factor of $O(k)$ (i.e.\ $H$ is a $O(k)$-spanner of $G$).
If $G$ is an unweighted graph, then $H$ has expected size $O(n^{1+1/k})$ and
is computed in $O(\log n\log^{*}n)$ depth with $O(m)$ work.
If $G$ is a weighted graph with ratio of maximum and minimum edge weights
bounded by $U$, $H$ has expected size $O(n^{1+1/k}\log k)$ and is computed
in $O(\log U\log n\log^{*}n)$ depth with $O(m)$ work.
\end{sloppypar}
\end{theorem}

Closely related to spanners are hopsets, which do not limit the edge count,
but aim to reduce the number of edges in the shortest paths.
These objects are crucial for speeding up parallel algorithms for (approximate)
shortest paths~\cite{KleinS97,Cohen00}.
Using the exponential start time clustering, we construct hopsets which lead to
the first parallel algorithm for approximating shortest paths in undirected
graphs with $O(m \poly\log n)$ work.
Here, our key idea is to employ backward analysis and analyze the algorithm
with respect to a single (unknown) optimal $s$-$t$ path.
We show that in expectation this path is not cut in too many places by
the decomposition scheme, and use this to bound the overall distortion.

\begin{theorem}
\label{thm:mainHopSet}
\begin{sloppypar}
There exists an algorithm that given as input an undirected, non-negatively 
weighted graph $G$, and parameters 
$\alpha,\epsilon\in(0,1)$,
preprocesses the graph in $O(m\epsilon^{-2 - \alpha}\log^{3 + \alpha}n)$
work and $O\left(n^{\frac{4 + \alpha}{4 + 2 \alpha} } \epsilon^{-1 -
\alpha}\log^2n\log^*n\right)$
depth, so that for
any vertices $s$ and $t$ one can return a $(1+\epsilon)$-approximation to the
$s-t$ shortest in $O(m \epsilon^{-1 - \alpha})$ work and
$O\left(n^{\frac{4 + \alpha}{4 + 2 \alpha} } \epsilon^{-2 - \alpha}\right)$ depth.
\end{sloppypar}
\end{theorem}

The paper is organized as follows: Section~\ref{sec:related} reviews some
standard notions and compares our results with related works.
In Section~\ref{sec:spanner}, we describe our spanner construction for both
unweighted and weighted graphs.
We then describe our hopset algorithm for unweighted graphs
in Section~\ref{sec:unweighted}, and extend it to the weighted setting in
Section~\ref{sec:weighted}.

Although the probabilistic analysis allows us to decouple some of the levels,
mild dependencies between the levels remain.
Such dependencies result in terms of $O(\log k)$ and $O(\log^2{n})$
in our results for spanners and hopsets respectively.
A promising direction for improvements is to construct
the clusterings on different levels in a dependent manner.
Picking the randomness across levels from the same source could
allow more streamlined analysis of the overall process.

\section{Background and Related Works}
\label{sec:related}

% Spanners and hopsets both aim to give succinct representations of
% distances in a graph.
We consider a graph $G=(V,E,w)$ with $\abs{V}=n$, $\abs{E}=m$ and edge
weights/lengths
$w:E\to\reals_+$.
Throughout the paper we will only deal with undirected graphs with
positive edge weights, so we can assume $w(e) \geq 1$ by normalizing and
$w(u,v)=w(v,u)$.
Furthermore, the graph is unweighted if $w(e)=1$ for all $e\in E$.
If $X$ is a subset of $V$ or $E$, we will use $G[X]$ to denote the induced
subgraph of $G$ on $X$.
% As we only work with shortest paths, we may assume that the graph is simple
% since multi-edges and self-loops can be removed.
If $H$ is a subgraph of $G$, we will use $G/H$ to denote the quotient graph
obtained from $G$ after contracting the connected components of $H$ into
points, removing self-loops and merging parallel edges (by keeping the shortest
edge).

% A decomposition of $G$ is a partition of $V$ into subsets $X_1,\cdots,X_k$.
% An edge $e$ is cut by such a decomposition if its two endpoints belongs to
% different $X_i$s.
% 
% All the objects that we construct aim to preserve distance in the original graph.
% We let $\dist_G(u,v)$ be the distance between $u$ and $v$ in $G$.
% As $G$ is undirected, $\dist_G(\cdot, \cdot)$ forms a metric.

The parallel performances of our algorithms are analyzed in the standard PRAM
model.
The longest sequence of dependent operations is known as depth,
while the total number of operations performed is termed work.
In practice, the abilities of algorithms to parallelize are often limited
by the number of processors.
For instance, a common assumption in the MapReduce model is that the number
of processors is $n^{\delta}$ for some small $\delta$~\cite{KarloffSS10}.
In such settings, an algorithm will fully parallelize as long as the depth is
less than $n^{1 - \delta}$.
As a result, it is more important to reduce work in order to obtain
speed-ups over sequential algorithms.

\subsection{Exponential Start Time Clustering}

\label{sec:ESTCluster}

We start by formalizing the key routine in this paper, a graph decomposition
routine which we call Exponential Start Time Clustering. %, see Figure.
It generates a partition of $V$ into subsets $X_1,\cdots,X_k$, and a center
$c_i$ for each $X_i$.
It also outputs a spanning tree for each cluster rooted at its center.
For convenience, %we will denote such partition as $\mathcal{X}$, and
if $v\in X_i$, we use $c(v)$ to denote $c_i$.
We use a routine from~\cite{MillerPX13}.

\begin{algo}[ht]
\textsc{ESTCluster}$(G,\beta)$
\vspace{0.05cm}

\underline{Input:}
Graph $G = (V, E, w)$, parameter $0 < \beta < 1$.

\underline{Output:} Decomposition of $G$.

\begin{algorithmic}[1]
	\STATE{For each vertex $u$, pick $\delta_u$ independently from the
    exponential distribution $\mathrm{Exp}(\beta)$.}
  \STATE{Create clusters by assigning each $v\in V$ to $u=\argmin_{u\in
  V}\{\dist(u, v) - \delta_u\}$, if $v=u$ we let it be the center its cluster.}
  \STATE{Return the clusters along with a spanning tree on each cluster rooted
  at its center}.
\end{algorithmic}

\caption{Exponential Start Time Clustering}
\label{alg:ESTCluster}
\end{algo}

In this paper we extend the algorithm to efficiently run on weighted graphs and 
also extend the analysis bounding the number of inter-cluster edges to more
general subgraphs.
The following lemma gives bounds on the run time and cluster diameter for the
weighted case.

%% Our applications uses additional properties of this routine,  which we extend to the
%% weighted setting.

\begin{lemma} (Theorem 1.2 from~\cite{MillerPX13})
  \label{lem:ESTCluster}
  Given a weighted graph $G=(V,E,w)$ where $\abs{V}=n$, $\abs{E}=m$,
  $w:E\to\mathbb{Z}_+$ with $\min_{e\in E}w(e)=1$,
  \textsc{ESTCluster}$(G,\beta)$ generates a set of disjoint clusters
  $\cup_iX_i=V$.
  % where each $X_i$ has diameter $O(\beta^{-1}\log{n})$ with high probability.
  The diameter of each $X_i$ is certified by a spanning tree on $X_i$, which
  has diameter at most $\frac k\beta\log n$ with probability at least
  $1-1/n^{k-1}$, for any $k\ge1$.
  This computation takes $O(\beta^{-1}\log{n} \log^{*} n)$ depth with high
  probability and $O(m)$ work.
\end{lemma}

We discuss the efficient implementation of the \textsc{ESTCluster} routine in
Appendix~\ref{sec:proofs}, as well as the effect different models of
parallelism have on the depth.
An analysis of the resulting decomposition in unweighted graphs is in Section 4
of~\cite{MillerPX13}, and it extends immediately to weighted graphs.
Our spanner algorithm requires a stronger variant of the edge cutting guarantee
which we state below and prove in Appendix~\ref{sec:proofs}.

Let $G$ be a weighted graph, a ball centered at $c$ of radius $r$ is defined as
$B(c,r)=\{v\in V\mid d(v,c)\le r\}$.
The center $c$ may either be a vertex or the midpoint of an edge.
% Given a decomposition $X_1 \ldots X_k$, we say a subset of the vertices $X$
% intersects $s$ clusters if it has non-empty intersections with $s$ or
% more of the $X_i$s.
We can show that balls with small radius do not intersect with too many
clusters.

\begin{lemma}
  \label{lem:intersectSubset}
  The probability that a ball of radius $r$ intersects $k$ or more clusters
  from \textsc{ESTCluster} is at most $\gamma^{k-1}$ where
  $\gamma=1-\exp(-2r\beta)$.
\end{lemma}

\begin{corollary}
  \label{cor:edgeCut}
  An edge $e$ with weight $w(e)$ is cut in the clustering produced by
  \textsc{ESTCluster} with probability at most $1-\exp(-\beta\cdot w(e))<\beta\cdot w(e)$.
\end{corollary}

\subsection{Spanners}

A subgraph $H$ of $G$ is said to be a $k$-spanner, if for every $u,v\in V$, we
have $\dist_{H}(u,v)\le k\cdot\dist_G(u,v)$, where $k$ is also called the
stretch factor.
Notice that it is sufficient to prove the stretch bound for endpoints of every
edge.
It is known that for any integer $k\ge1$, any undirected graph with $n$
vertices admits $(2k-1)$-spanners with $O(n^{1+1/k})$ edges\footnote{Note that
$O(n^{1+1/k})=O(n)$ for $k=\Omega(\log n)$.}, and this is essentially the best
tradeoff between sparsity and stretch \cite{PelegS89,ThorupZ05}.

\begin{figure*}[ht!]

\begin{center}
\textbf{weighted graphs}

\begin{tabular}{ccccc}
multiplicative distortion & Size & Work & Parallel depth & Notes\\
\hline
$2k - 1$ & $\frac{1}{2} n^{1 + 1/k}$ & $O(mn^{1 + 1/k})$ & $O(n^{1 + 1/k})$ & \cite{AlthoferDDJS93}\\
$2k - 1$ & $O(kn^{1 + 1/k})$ & $O(km)$ & $O(k\log^*n)$ & \cite{BaswanaS07}\\
% $2k - 1$ & $O(kn^{1 + 1/k})$ & $O(km)$ & $O(\log{n})$ & \cite{RodittyTZ05}\\
$O(k)$ & $O(n^{1+1/k}\log k)$ & $O(m)$ & $O(k\log^{*}{n}\log{U})$ & new
\end{tabular}

\vspace{0.1in}

\textbf{unweighted graphs}

\begin{tabular}{ccccc}
multiplicative distortion & Size & Work & Parallel depth & Notes\\
\hline
% $2k - 1$  & $n^{1 + 1/k}$ & $O(m)$ & $O(\log^3{n})$ & \cite{Peleg00}\\
$2k - 1$ & $O(k n^{1 + 1/k})$ & $O(km)$ & $O(k\log^*n)$ & \cite{BaswanaKMP10}\\
$O(2^{\log^{*}{n}}\log n)$ & $O(n)$ & $O(m \log{n})$ & $O(\log{n}\log^*n)$ & \cite{Pettie08}\\
$O(k)$ & $O(n^{1+1/k})$ & $O(m)$ & $O(k\log^{*}{n})$ & new
\end{tabular}

\caption{Known results for spanners, $U$ represents the range of weights.}
\label{fig:spannerList}

\end{center}

\end{figure*}

A summary of parallel algorithms for constructing spanners can be found in
Figure~\ref{fig:spannerList}.
Our algorithms improve upon the $O(k)$ overhead in spanner sizes from previous
parallel algorithms while losing constant factors in the stretch.
On unweighted graphs, this improvement comes mainly from the ability
to invoke exponential start time decomposition in the spanner
construction by Peleg and Schaffer algorithm~\cite{PelegS89}.
Our extension of this routine to the weighted case relies on the probabilistic
aspects of the decomposition.
This leads to improvements by factors of $\frac{k}{\log k}$ in spanner size and
factors of $k$ in work over the previous best~\cite{BaswanaS07}.
Such routines are also directly applicable to the graph sparsification
algorithm by Koutis~\cite{Koutis14}.

Spanners have also been studied under additive error
\cite{DorHZ00,BaswanaKMP05}.
An $(\alpha,\beta)$-spanner is a subgraph that preserves distances up to a
multiplicative factor $\alpha$ and an additive term $\beta$.
In the table above, the construction by \cite{BaswanaKMP10} in fact produces
$(k,k-1)$-spanners in unweighted graphs.
For a single edge, this gives a multiplicative stretch factor of at most
$2k-1$, which we listed for comparison purposes.

This problem has also been extensively studied in the distributed
setting~\cite{BaswanaS07,DerbelGP07,DerbelGPV08,Pettie08}.
In the table above, the constructions by~\cite{BaswanaS07}, \cite{Pettie08} and
\cite{BaswanaKMP10} can produce spanners of the same quality in the
synchronized distributed model, where the number of rounds of the algorithm is
given by the stretch factor and unit size messages are used.
Our spanner construction for unweighted graphs can also be ported to this
distributed setting with similar guarantees, as its employs breadth first
search, which admits a simple implementation in synchronized distributed
networks.
However for finding spanners in weighted graphs, it is necessary for us to
contract parts of the original graph and work on the resulting quotient graph.
Thus we lose the ability apply it to the distributed setting with weighted
edges.

\subsection{Hopsets}

Hopsets were formalized by Cohen~\cite{Cohen00} as a crucial component
for parallel shortest path algorithms.
The goal is to add a set of extra edges to the graph
so that the $h$-hop distance in the new graph suffices for a good approximation.
Let weight of a path $p$, denoted as $w(p)$, be the sum of weights of all edges
on it, $w(p)=\sum_{e\in p}w(e)$.
The distance between two vertices $s$ and $t$, $\dist(s, t)$,
is the weight of the shortest (lightest) $s$-$t$ path.
Furthermore, with a set of edges $E'$, the $h$-hop distance between $s$ and $t$
in $E'$, denoted by $\dist_{E'}^h (u,v)$, is defined to be the weight of the
minimum weight path with at most $h$ edges between $s$ and $t$, using only
edges from $E'$.
If $h$ is omitted we assume $h=n$, if $E'$ is omitted we assume $E'=E$.
A probabilistic version of hopsets can be described as follows:

\begin{definition}
\label{def:hopset}
Given a graph $G = (V, E, w)$,
a $(\epsilon, h, m')$-hopset is a set of edges $E'$ such that:
\begin{enumerate}
\item $|E'| \leq m'$.
\item Each edge $uv$ in $E'$ corresponds to a $uv$-path $p$ in $G$ such that
$w(uv) = w(p)$.
\item For any vertices $u$ and $v$, with probability $1/2$ we have:
\begin{align*}
\dist_{E  \cup E'}^h (u,v) \leq (1 + \epsilon) \dist_{E} (u, v).
\end{align*}
\end{enumerate}
\end{definition}

Given such a hopset, a result by Klein and and Subramanian~\cite{KleinS97} allows
us to approximate the length of the path efficiently.
They showed that when given an  $(\epsilon, h, m')$-hopset, a shortest path  
can be found in $O(m \epsilon^{-1})$ work and $O(hn^{\alpha}\epsilon^{-1})$
depth.
As a result, the main work in parallel shortest path algorithms is to
efficiently compute hopsets.
A summary of previous algorithms, as well as ours, is below in
Figure~\ref{fig:hopsets}.

\begin{figure*}[ht!]

\begin{center}

\begin{tabular}{ccccc}
Hop count & Size &  Work &  Depth & Notes\\
\hline
$O(n^{1/2})$ & $ O(n) $ & $O(mn^{0.5})$ & $O(n^{0.5}\log{n})$ & \cite{KleinS97,ShiS99}, exact\\

$O(\poly\log n)$ & $O(n^{1+\alpha}\poly\log n)$ & $\tilde O(mn^{\alpha})$ & $O(\poly{\log{n}})$  & \cite{Cohen00}\\

$(\log n)^{O((\log\log n)^2)}$ & $O\left(n^{1+O(\frac{1}{\log\log n})}\right)$ & $\tilde
O\left(mn^{O(\frac{1}{\log\log n})}\right)$ & $(\log n)^{O((\log\log n)^2)}$  & \cite{Cohen00}\\

$O(n^{\frac{4 + \alpha}{4 + 2 \alpha} })$ &  %\epsilon^{-2 - \alpha}
$O(n )$ & %\epsilon^{-1 - \alpha}
$O(m \log^{3 + \alpha} n)$ & %\epsilon^{-2 - \alpha}
$O(n^{\frac{4 + \alpha}{4 + 2 \alpha} } )$ & new %\epsilon^{-1 - \alpha} )
\end{tabular}

\caption{Performances of Hopset Constructions,
omitting $\epsilon$ dependency.}
\label{fig:hopsets}

\end{center}

\end{figure*}

For Cohen's algorithm, $\Omega(n^{\alpha})$ processors
are needed for parallel speedups in both the construction and query stages
\footnote{A more detailed analysis leads
to a tighter bound of $\Omega(\exp(\sqrt{\log{n}}))$}.
In our case, if $\epsilon$ is a constant, $O(\log^{3 + \alpha}n)$
processors are sufficient for parallel speedups.
Furthermore, once a hopset is constructed, even a constant number
of processors suffices for speedups.
% 
% We will assume that the edge weights are polynomially bounded
% due to existing reductions such as the one by Klein and Sairam~\cite{KleinS92}.
% The reduction we use can be found be found in the full version of this paper
% \cite{MPVX:arxiv}.

\section{Spanners}

\label{sec:spanner}

We first describe our spanner construction.
Our spanner construction in unweighted graphs has the same structure as the
sequential routine by Peleg and Schaffer~\cite{PelegS89}:
after the decomposition step, we add in single edges between adjacent clusters.
%The number of extra edges added then follows from
%Corollary~\ref{cor:few-neighbor-clusters}.
We formalize this algorithm for completeness here.
\begin{algo}[ht]
 \textsc{UnweightedSpanner}$(G,\delta)$
 \vspace{0.05cm}

 \underline{Input:} 
 An unweighted graph $G$ and parameter $k\ge1$.
 \underline{Output:}
 A $O(k)$-spanner of $G$.

 \begin{algorithmic}[1]
   \STATE Compute an exponential start time clustering with $\beta=\frac{\log n}{2k}$,
   let $H$ be the forest produced.
   \STATE From each boundary vertex, add to $H$ one edge connecting to each adjacent
   cluster.
   \STATE Return $H$.
 \end{algorithmic}

 \caption{Spanner construction for unweighted graphs.}
 \label{alg:unweighted-spanner}
\end{algo}

The Peleg and Schaffer algorithm~\cite{PelegS89} relied on a bound introduced
by Awerbuch~\cite{Awerbuch85}, which bounds the number of interacting clusters
around a single vertex.
The same bound can be obtained with exponential start time clusterings using
Lemma~\ref{lem:intersectSubset}.

\begin{corollary}
  \label{cor:few-neighbor-clusters}
  In an exponential start time decomposition with parameter
  $\beta=\frac{\log n}{2k}$, for any vertex $v\in V$, the ball
  $B(v,1)=\{u\in V\mid d(u,v)\le 1\}$ intersects $O(n^{1/k})$ clusters in
  expectation.
\end{corollary}

\Proof
By Lemma~\ref{lem:intersectSubset}, $B(v,1)$ intersects $k$ or more clusters
with probability at most $(1-\exp(2\beta))^{k-1}$.
Let $L$ be the number of clusters intersecting $B(v,1)$, we then have
\begin{align*}
  \expct{L}
  =
  \sum_{l=1}^\infty\prob{L\ge l}
  &\le
  \sum_{l=1}^\infty(1-\exp(-2\beta))^{l-1}
  \\&=
  \frac1{\exp(-2\beta)}
  \\&=
  \frac1{\exp(-\log n/k)}
  \\&=
  n^{1/k}.
\end{align*}
\QED

\begin{lemma}
  \label{lem:unweighted-spanner}
  Given a connected unweighted graph and for any $k\ge1$,
  \textsc{UnweightedSpanner} constructs a $O(k)$-spanner with high probability
  of expected size $O(n^{1+1/k})$.
  This takes $O(k\log^*n)$ depth with high probability and $O(m)$
  work.
\end{lemma}

\Proof
The algorithm starts by constructing an exponential start time decomposition
with parameter $\beta=\frac{\log n}{2k}$.
Let $F$ be the forest obtained from the decomposition, notice that $F$ has at
most $n-1$ edges.
Then for each boundary vertex $v$, (i.e.\ $v$ is incident to an inter-cluster
edge), we add one edge between $v$ and each of the adjacent clusters to our
spanner.
Using Corollary~\ref{cor:few-neighbor-clusters} and considering the ball
$B(v,1)$ for each $v\in V$, we see that $O(n^{1+1/k})$ edges are added this way
in expectation.

For an edge $e$ internal to a cluster, its stretch is certified by the spanning
tree within the cluster, whose diameter is bounded by $O(k)$ with high
probability by Lemma~\ref{lem:ESTCluster}.
If the edge $e$ has its endpoints in two different clusters, our spanner must
contain some edge between these two clusters. As with high probability both of
these clusters has diameter $O(k)$, the stretch of $e$ is agin bounded
by $O(k)$ with high probability.
% Notice that singleton vertices do not contribute any edges to our spanner, thus
% should not be counted.
The depth and work bounds also follow from Lemma~\ref{lem:ESTCluster}.
\hspace*{\fill}\QED

This routine can be extended to the weighted setting by bucketing
the edges by powers of $2$.
Given $G=(V,E,w)$ where $U=(\max_ew(e))/(\min_ew(e))$ is the ratio
between maximum and minimum edge weights, we bucket the edges as
\begin{gather*}
  E_i=\{e\in E\mid w(e)\in[2^{i-1},2^i)\}.
\end{gather*}
This allows us to run the unweighted algorithm on essentially disjoint sets of
edges, but leads to an overhead of $O(\log{U})$ in the total size.
We reduce this overhead to $O(\log k)$ using an approach introduced
in~\cite{CohenMPPX14:arxiv} that's closely related to the AKPW low-stretch
spanning tree algorithm~\cite{AlonKPW95}.
We build spanners on these buckets in order, but contract the low-diameter
components with smaller weights.
Lemma~\ref{lem:unweighted-spanner} then allows us to bound the expected
rate at which vertices are contracted, and in turn the size of the spanner.

Our contraction scheme is significantly simpler than previous ones because
we will be able to ensure that edge weights in different levels differ by
factors of $\poly{k}$, where $k$ is the stretch factor.
To this end, we first break up the input graph into $O(\log k)$ graphs
where edge lengths are well separated.
We define $G_j$ to be the graph with vertex set $V$ and edge set
$\cup_{i\ge0}E_{j+i\cdot c\lg k}$, where $c$ is an appropriate
constant\footnote{The constant $c$ should be chosen to achieve the desired
succes probability from Lemma~\ref{lem:ESTCluster}. We will hide $c$ inside
big-O notations from now on.}.
It is clear that the union of $O(\log k)$ such $G_j$s form the whole graph, and
they all have $O(\log U)$ buckets of edges, where weights differ by at least
$O(k^c)$ between different buckets.
Thus if we can find a $O(n^{1+1/k})$-sized spanner for each of $G_j$, we
obtain a $O(n^{1+1/k}\log k)$-sized spanner for $G$.

For each $G_j$, we use the fact that the weights are well-separated to form
hierarchical contraction schemes.
Pseudocode of this algorithm is given in Algorithm~\ref{alg:wellSeparatedSpanner}.
%%Here we use a hierarchical graph contraction scheme first introduced in
%%\cite{AlonKPW95}
%and the form in which it appeared in \cite{CohenMPPX13}.
%%As we interact with of this algorithm rather closely, we describe the
%%contraction along with our spanner construction it in Algorithm~\ref{alg:wellSeparatedSpanner}.

\begin{algo}[h]
  \textsc{WellSeparatedSpanner}$(G)$
  \vspace{0.05cm}

  \underline{Input:} 
  A weighted graph $G$ with well separated edge weight buckets as described above.

  \underline{Output:}
  A $O(n)$-sized spanner for $G$.

  \begin{algorithmic}[1]
    \STATE Relabel the edge buckets to be $A_1,A_2,\dots,A_s$ in increasing order of
      weights, such that edges in $A_i$ have weights in $[w_i,2w_i)$ and
      $w_{i+1}/w_i\ge O(k)$.
    \STATE Initialize $H_0=\emptyset$ and $S=\emptyset$.
    \FOR{$i=1$ to $s$}
      \STATE Let $\Gamma_i=G[A_i]/H_{i-1}$ with uniform edge weights.
      \STATE Perform \textsc{ESTCluster} with $\beta=\frac{\log n}{2k}$ on $\Gamma_i$
      \STATE Let $F$ be the forest produced in the previous step.
      \STATE $S=S\cup F$ and $H_i=H_{i-1}\cup F$.
      \STATE For each boundary vertex, add one edge connecting each of the
      adjacent clusters to $S$. \label{ln:augment}
    \ENDFOR
    \RETURN $S$.
  \end{algorithmic}

  \caption{Spanner Construction on graphs with well separated edge weights.}
  \label{alg:wellSeparatedSpanner}
\end{algo}

\begin{theorem}
  \label{thm:weighted-spanner}
  Given a weighted graph $G$ with $n$ vertices, $m$ edges and for any $k\ge1$,
  we can compute with hight probability a $O(k)$-spanner for $G$ of expected
  size $O(n^{1+1/k}\log k)$, using $O(k\log^*n\log U)$ depth and $O(m)$
  work.
\end{theorem}

\Proof
As discussed above, we break $G$ into $O(\log k)$ edge-disjoint graphs in which
edge weights are well separated.
We run \textsc{WellSeparatedSpanner} on each of these graphs in parallel, each
iterations of the loop performs an exponential start time decomposition on
disjoint sets of edges, thus the overall work is $O(m)$.
As there are $O(\log U)$ iterations, the overall depth is
$O(k\log^*n\log U)$ with high probability.

Now we show that \textsc{WellSeparatedSpanner} produces a spanner for each of
these graphs.
In each iteration of the for-loop, the unweighted algorithm is run on
$\Gamma_i=G[A_i]/H_{i-1}$.
This produces an unweighted spanner for edges in $A_i$ by
Lemma~\ref{lem:unweighted-spanner}.
Since the edge weights differ by at least $O(k)$ between levels, using
Lemma~\ref{lem:ESTCluster} and induction on the loop index we see
that vertices in the quotient graph $\Gamma_i$ corresponds to pieces of
diameter at most $w_i$ in the spanner constructed so far, with high
probability.
Therefore the stretch bound for edges from $A_i$ in $\Gamma_i$ gets worse by at
most a factor of $2$ when translated in $G$.

We finish by bounding the size of our spanner.
% which amounts to bounding the
% total number of vertices with degree at least one in all of $\Gamma_i$s, as
% singleton vertices do not contribute to the size bound from
% Lemma~\ref{lem:unweighted-spanner}.
Using an argument similar to the proof of Lemma~\ref{lem:ESTCluster}, we notice
that any non-singleton vertex in one of $\Gamma_i$ has probability at
least $1/n^{1/k}$ of being contracted away.
Thus in expectation each vertex participates in $O(n^{1/k})$ level of
\textsc{WellSeparatedSpanner}, and in each level it contributes $O(n^{1/k})$
inter-cluster edges and $O(1)$ forest edges in expectation.
Thus \textsc{WellSeparatedSpanner} produces a spanner of size $O(n^{1+2/k})$,
where the exponent $1+2/k$ can be reduced to $1+1/k$ if we back down on the
stretch bound by a factor of $2$.
Since the graph is decomposed into $O(\log k)$ well separated graphs, this
gives us the $O(n^{1+1/k}\log k)$ overall bound on the expected size of our
spanner.
\QED

\section{Hopsets in Unweighted Graphs}

\label{sec:unweighted}

\begin{figure*}[ht!]
\begin{center}

\includegraphics{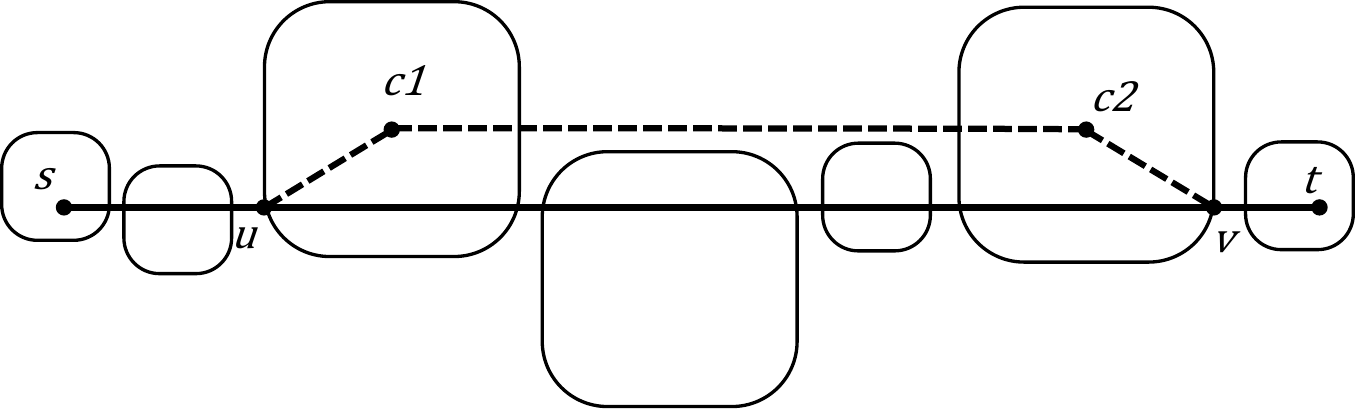}
% \begin{tikzpicture}[x=1cm, y=1cm, scale=0.75]
% 
% \tikzstyle{every node}=[draw,fill,shape=circle,inner sep=0,minimum size=0.2cm]
% \coordinate (s) at (1,0);
% \coordinate (t) at (15,0);
% \coordinate (c1) at (5, 0.8);
% \coordinate (c2) at (12.55, 0.7);
% \coordinate (u) at (3.4, 0);
% \coordinate (v) at (14.2, 0);
% \node(S)[label = $s$] at (s) {$$};
% \node(T)[label = $t$] at (t) {$$};
% \draw[ultra thick] (s) -- (t);
% \draw(0.8, 0.5) circle (0.8);
% \draw(2.4, -0.2) circle (0.8);
% \draw(15.2, -0.3) circle (0.8);
% \node(C1)[label = $c1$] at (c1) {$$};
% \node(C2)[label = $c2$] at (c2) {$$};
% \node(U)[label = $u$] at (u) {$$};
% \node(V)[label = $v$] at (v) {$$};
% \draw(c1) circle (1.8);
% \draw(c2) circle (1.8);
% \draw[ultra thick, dotted] (u)--(c1)--(c2)--(v);
% \draw(7.9, -1) circle (1.5);
% \draw(9.9, 0.4) circle (0.8);
% \end{tikzpicture}

\caption{Interaction of an $s-t$ path with the decomposition scheme.
Shortcut edges connecting the centers of large clusters allow us to `jump'
from the first vertex of $p^*$ in a large cluster ($u$), to the last vertex of $p^*$ in a large cluster ($v$).
The edges $uc_1$, $c_2v$ are star edges, while $c_1c_2$ is a clique edge.}
\label{fig:pathcut}
\end{center}

\end{figure*}

Our hop-set construction is based on recursive application of the exponential
start time clustering from Section~\ref{sec:ESTCluster}.
We will designate some of the clusters produced,
specifically the larger ones, as special.
Since each vertex belongs to at most one cluster, there cannot be too many
large clusters.
As a result we can afford to compute distances from their centers to
all other vertices, and keep a subset of them as hopset edges in the graph.
There are two kinds of edges that we keep:
\begin{enumerate}
  \item \emph{star edges} between the center of a large cluster and all
    vertices in that cluster.
	\item \emph{clique edges} between the center of a large cluster and the
    centers of all other large clusters.
\end{enumerate}
In other words, in building the hopset we put a star on top of each large
cluster and connect their centers into a clique.
Then if our optimal $s$-$t$ path $p^{*}$ encounters two or more of these large
clusters, we can jump from the first to the last by going through their
centers.
One possible interaction between the decomposition scheme and a path $p^{*}$
in one level of the algorithm is shown in Figure~\ref{fig:pathcut}.

This allows us to replace what hopefully is a large part of $p^*$ with only
three edges: two star edges and one clique edge.
%two between centers and vertices on $p$, and one between the centers.
However this replacement may increase the length of the path by the
diameter of the large clusters.
But as this distortion can only happen once, it is acceptable as long as the
diameter of the clusters are less than $\epsilon w(p^*)$.
Our algorithm then recursively builds hopsets on the small clusters.
The probabilistic guarantees of an edge being cut gives that
$p^*$ does not interact with too many such clusters.
So once again a reasonable distortion within these clusters can be incurred.

Formally, two parameters are crucial to our algorithm:
the parameter $\beta$ with which the decomposition routine is run, and
the threshold $\rho$ by which a cluster is deemed large.
The algorithm then has the following main steps:
\begin{enumerate}

\item Compute a  exponential start time clustering with parameter $\beta$

\item Identify clusters with more than $n/\rho$ vertices as large clusters.

\item Construct star and clique edges from the centers of each large cluster.

\item Recurse on the small clusters.

\end{enumerate}

Our choice of $\beta$ at each level of the recursion is constrained
by the additive distortion that we can incur.
%Consider recursing on a ball that contains a path $p$ of length $d$.
%The best case scenario in terms of hop-count is that a shortcut
%involving the center of two large clusters is taken.
%However, this detour costs us an additive distortion proportional
%to the diameter of the clusters.
%As a result, the diameter of the smaller pieces needs to be
%comparable to $\epsilon$ times the expected length of the intersection of $p$
%with the small pieces.
%
%In terms of the $\beta$s,
Consider a cluster obtained at
the $i\textsuperscript{th}$ level of the decomposition ran with $\beta_i$.
Since the path has length $d$ and each edge is cut with probability $\beta_i$,
the path is expected to be broken into $\beta_i d$ pieces.
Therefore on average, the length of each piece in a cluster
is about $\beta_i^{-1}$.
The diameter of a cluster in the next level on the other hand
can be bounded by $k \beta_{i + 1}^{-1} \log{n}$, where the constant $k\ge1$
can be chosen with desired success probability using
Lemma~\ref{lem:ESTCluster}.
Therefore, we need to set $\beta_{i + 1}$ so that:
\begin{align*}
k \beta_{i + 1}^{-1} \log{n}
& \leq \epsilon \beta_i^{-1}\\
\beta_{i + 1}
& \geq \left( k\epsilon^{-1} \log{n}  \right) \beta_i.
\end{align*}
In other words, the $\beta$s need to increase from one level to the next
by a factor of $k\epsilon^{-1} \log{n} $ where $\epsilon<1$ the
distortion parameter.
This means that the path $p^*$ is cut with granularity that
increases by a factor of $O(\epsilon^{-1}\log{n})$ each time.
Note that the number of edges cut in all level of the recursion serves as
a rough estimate to the number of hops in our shortcut path.
Therefore, a different termination condition is required to ensure
that the path is not completely shattered by the decomposition scheme.
As we only recurse on small clusters, if we require their size to decrease at a
much faster rate than the increase in $\beta$, our recursion will terminate
with most pieces of the path within large clusters.
To do this, we introduce a parameter $\rho$ to control this rate of
decrease.
Given a cluster with $n$ vertices, we designate a cluster
$X_i$ to be small if $\abs{X_i} \leq n/\rho$.
As our goal is a faster rate of decrease, we will set
$\rho =\left( k\epsilon^{-1}\log n \right)^{\delta}$
%%\begin{align*}
%%  \rho =\left(\frac{\log n}{\epsilon} \right)^{\delta}
%%\end{align*}
for some $\delta > 1$.

Pseudocode of our hopset construction algorithm
is given in Algorithm~\ref{alg:hopSet}.
Two additional parameters are needed to control the first and
last level of the recursion:
$\beta=\beta_0$ is the decomposition parameter on the top level,
and $n_{\mit{final}}$ is the base case size at which the recursion stops.

\begin{algo}[ht]
  \textsc{HopSet}$(V,E,\beta)$.
  \vspace{0.05cm}

  \underline{Input:} 
  Undirected, unweighted graph $G = (V, E)$ and decomposition parameter
  $\beta$.
  %size reduction parameter $\rho = \beta^{\delta}$.

  \underline{Output:}
  The input graph $G$ augmented with a set of weighted edge $E^*$.

  \begin{algorithmic}[1]
    \STATE If $\abs{V}\le n_{\mathit{final}}$, exit.
%    \STATE Let $\beta=\left( c \log n \epsilon^{-1} \right)B$ where $c$ is the
%    constant from \textsc{ESTCluster}.
%%    \STATE Let $\beta=\left(\frac{4c\log n}\epsilon\right)B$ where $c$ is the
%%    constant  from \textsc{ESTCluster}.
    \STATE Compute a exponential start time clustering of $G$ with parameter $\beta$,  $\mathcal{X}$.
    \label{algline:ESTCluster}
    \IF{this is first call}
    \STATE For each cluster $X$ and \emph{in parallel},  recursively call
    \textsc{HopSet}$(X, E(X), ( k \epsilon^{-1}\log n ) \beta)$.
    \ELSE
    \STATE Let $\mathcal{X}_b=\{X\in\mathcal{X}\ :\ \abs{X}\ge\abs{V}/\rho\}$
    be the set of large clusters.
    \STATE Let $\mathcal{X}_s=\{X\in\mathcal{X}\ :\ \abs{X}<\abs{V}/\rho\}$ be
    the set of small clusters.
    \STATE For each large cluster $X\in\mathcal{X}_b$ with center $c$ and
    $v\in X$, add a star edge $(v,c)$ with weight $\dist(v,c)$
    \label{algline:makeStar}
    \STATE For all pairs of large clusters $X_1,X_2\in\mathcal{X}_b$ with
    centers $c_1$, $c_2$ respectively, add a clique edge $(c_1,c_2)$ with
    weight $\dist(c_1,c_2)$.
    \label{algline:makeClique}
    \STATE For each $X\in\mathcal{X}_s$, recursively call
    \textsc{HopSet}$(X, E(X), ( k \epsilon^{-1}\log n ) \beta)$ \emph{in parallel}.
    \ENDIF
  \end{algorithmic}

  \caption{Hop Set Construction on Unweighted Graphs}
  \label{alg:hopSet}
\end{algo}

We start with the following simple claim about the $\beta$
parameters in the recursion.

\begin{claim}
  \label{claim:UWHSparemeters}
  If the top level of the recursion is called with $\beta=\beta_0$ as the input parameter.
  then the parameter $\beta$ in $i$\textsuperscript{th}a level,
  denoted $\beta_i$, is given by $\beta_i=\left( k \epsilon^{-1}\log{n}
  \right)^{i}\beta_0$.
%%\end{align*}
%%\begin{align*}
%%    \beta_i=\left(\frac{4c\log n}\epsilon\right)^i\beta_0
%%\end{align*}
\end{claim}

We now describe how hopsets are used to speed up the parallel BFS.
We prove the lemma in the generalized weighted setting as it will become useful in Section
\ref{sec:weighted}.

%  $E' = \textsc{HopSet}(V, E, B, D)$ is a
%  $(\epsilon, \log_{K}n, n^{1/\alpha}BD\dist_E(u,v), n^{1-\gamma}\log^{2\alpha}n)$
%  -hopset.
\begin{lemma}
  \label{lem:hopAndDistortion}
  Given a weighted graph $G=(V,E,w)$ with $\abs{V}=n$ and $\abs{E}=m$,
  let $E'$ be the set of edges added by running
  \textsc{HopSet}$(V,E,\beta_0)$.
  Then for any $u,v\in V$, we have with probability at least $1/2$:
  \begin{align*}
    \dist_{E\cup E'}^h(u,v)\le\dist_E(u,v)+O(\epsilon\log_\rho n\cdot\dist_E(u,v))
  \end{align*}
  where $h=n_{\mit{final}}^{1-1/\delta}n^{1/\delta}\beta_0\dist_E(u,v)$.
\end{lemma}

\Proof
Let $p$ be any shortest path with endpoints $u$ and $v$, we show how to
transform it into a path $p'$ satisfying the above requirements using edges in
$E'$.
In each level of the algorithm, the clustering routine breaks $p$ up
into smaller pieces by cutting some edges of $p$.
Consider an input subgraph in the recursion that intersects the path $p$
from vertex $x$ to vertex $y$.
The decomposition partitions this intersection into a number of segments,
each contained in a cluster.
Starting from $x$, we can identify the first segment that is contained in a
large cluster, whose start point is denoted by $x'$, and similarly we can find
the last segment contained in a large cluster with its end point denoted by
$y'$.
We drop all edges on $p$ between $x'$ and $y'$ and reconnect them using three
edges $(x',c(x'))$, $(c(x'),c(y'))$ and $(c(y'),y')$.
%by jumping from $x'$
%to the center of the cluster containing $u'$ using a star edge, from there we
%then jump to the center of the cluster containing $v'$ via a clique edge,
%followed by a final jump to $v'$ via a star edge.
We will refer to this procedure as shortcutting.
We then recursively build the shortcuts on each segment before $u'$ and after $v'$.
Note that these segments are all contained in small clusters, thus they are
also recursed on during the hopset construction.
We stop at the base case of our hopset algorithm.

We first analyze the number of edges in the final path $p'$, obtained by
replacing some portion of $p$ with shortcut edges.
The path $p'$ consists of edges cut by the decomposition, shortcut edges that
we introduced, and segments that are contained in base case pieces.
It suffices to bound the number of cut edges, as the segments in $p'$ separated
by the cut edges have size at most the size of the base case.
Recall from Lemma \ref{lem:intersectSubset} that any edge of weight
$w(e)$ has probability $\beta w(e)$ of being cut in the clustering.
Thus, the expected number of cut edges can be bounded by
\begin{align*}
  \sum_{e\in p}\left(\sum_i\beta_i\right)w(e)
&= \left(\sum_i\beta_i\right)w(p).
\end{align*}
Since $\beta_i$s are geometrically increasing, we can use the approximation
$\sum_i\beta_i\approx\beta_l$, where $l=\log_\rho n$ is the depth of
recursion.
Recalling that $\rho=(k\epsilon^{-1}\log n)^\delta$:
\begin{align*}
  \beta_ld
  &=
  \left(k\epsilon^{-1}\log
  n\right)^{\log_\rho\left(\frac{n}{n_{\mit{final}}}\right)} \beta_0d
  \\&=
  \left(\rho^{1/\delta}\right)^{\log_\rho\left(\frac{n}{n_{\mit{final}}}\right)}\beta_0d
  \\&=
  \left(  \frac{n}{n_{\mit{final}}} \right)^{1/\delta}\beta_0d.
\end{align*}
As the recursion terminates when clusters have fewer than $n_{\mit{final}}$
vertices, each path in such a cluster can have at most $n_{\mit{final}}$ hops.
Multiplying in this factor gives $n^{1/\delta} n_{\mit{final}}^{1 - 1 / \delta} \beta_0d$.

Next we analyze the distortion introduced by $p'$ compared to the original path
$p$.
% The set of cut edges are shared by both $p$ and $p'$ so no distortion is
% introduced.
Shortcutting in level $i$ introduces an additive distortion of at most $4c
\beta_i^{-1} \log n $.
The expected number of shortcut made in level $i$, in other words the expected
number of cluster in $(i-1)$\textsuperscript{th} level intersecting the path
$p$, is bounded by $\beta_{i-1}d$.
Thus the amount of additive distortion introduced in level $i$ is at most
\begin{align*}
  \left( \beta_{i-1}d \right) \cdot\frac{4c\log n}{\beta_i}
& = O(\epsilon d).
\end{align*}
This gives an overall additive distortion of $O(\epsilon d\log_\rho n)$.
\QED

\begin{lemma}
  \label{lem:hopSetSize}
  If \textsc{HopSet} is run on a graph $G$ with $n$ vertices, it adds at most
  $n$ star edges and $O\left((n/n_{\mit{final}})\log^{2\delta}n\right)$ clique edges
  to $G$.
\end{lemma}

\Proof
As we do not recurse on large clusters, each vertex is part of a large
cluster at most once.
As a result, we add at most $n$ edges as star edges in Line
\ref{algline:makeStar} of \textsc{HopSet}.

To bound the number of clique edges, we claim that the worst case is when we
always generate small clusters, except in the level above the base cases, where
all the clusters are large.
Suppose an adversary trying to maximize the number of clique edges decides
which clusters are large.
Since we do not recurse on large clusters, if on any level above the base case we
have a large cluster, the adversary can always replace it with a small cluster,
losing at most $\rho$ clique edges doing so (since there are at most
$\rho$ large clusters), and gain $\rho^2$ edges in the next
level by making the algorithm recurse on that cluster.
Since the base case clusters have size at most $n_{\mit{final}}$, there are at most
$n/n_{\mit{final}}$ clusters in the level above, where each cluster adding at most
$\rho^2$ edges.
Therefore at most
$(n/n_{\mit{final}})\rho^2=(n/n_{\mit{final}})(\log n/\epsilon)^{2\delta}$ edges are
added in total.
\QED

\begin{theorem}
  \label{thm:unweightedHopSet}
  Given constants $\delta>1$ and $\gamma_1<\gamma_2<1$, we can construct a
  $(\epsilon \log{n},h,O(n))$-hopset on a graph with $n$ vertices and $m$ edges in
  $O( n^{\gamma_2}\log^2n\log^*n)$ depth and
  $O(m\log^{1+\delta}n \epsilon^{-\delta})$
  work, where $h=n^{1+1/\delta+\gamma_1 \left(1 - 1 / \delta \right) -\gamma_2}$.
\end{theorem}

\Proof
We claim that the theorem statement can be obtained by setting
$\beta_0=n^{-\gamma_2}$ and $n_{\mit{final}}=n^{\gamma_1}$.
The correctness of the constructed hopset follows directly from Lemma
\ref{lem:hopAndDistortion}, Lemma \ref{lem:hopSetSize}, and the fact that any
path in an unweighted graph has weight at most $n$.
Specifically, for any vertices $u$ and $v$ with $\dist(u,v)=d$, the expected
hop-count is:
\begin{align*}
  n^{1/\delta} n_{\mit{final}}^{1 - 1 / \delta} \beta_0 d
& \leq n^{1/\delta} n^{\gamma_1 \left(1 - 1 / \delta_2 \right)}
	n^{- \gamma_2} n
\\& =
n^{1 + 1 / \delta + \gamma_1 \left(1 - 1 / \delta \right) - \gamma_2}
\end{align*}
and the expected distortion is $O(\epsilon \log n\cdot d)$.
By Markov's inequality, the probability of both of these exceeding four
times their expected value is at most $1/2$, and the result can be
obtained by adjusting the constants.

So we focus on bounding the depth and work.
As the size of each cluster decreases by a factor of $\rho$ from one level
to the next, the number of recursion levels is bounded by
$\log_\rho (n/ n_{\mit{final}})$.
As $n / n_{\mit{final}}$ is polynomial in $n$ with our choice of parameters,
we will treat this term as $\log{n}$.

% We start with the decomposition schemes called by
The algorithm starts by calling \textsc{HopSet}$(V,E,n^{-\gamma_2})$.
Since the recursive calls are done in parallel, it suffices to bound the time
spent in a single call on each level.
Lemma \ref{lem:ESTCluster} gives that the clustering takes
$O(\beta^{-1} \log n\log^*n)$ depth and linear work.
Since the value of $\beta$ only increases in subsequent levels,
all decompositions in each level of the recursion
can be computed in $O(n^{\gamma_2}\log n\log^*n)$ depth and $O(m)$ work.
This gives a total of $O(n^{\gamma_2}\log^2n\log^*n)$ depth
and $O(m \log{n})$ work from Line \ref{algline:ESTCluster}.
In addition, Line \ref{algline:makeStar} can be easily incorporated into the
decomposition routine at no extra cost.

To compute the all-pair shortest distances between the centers of the large
clusters (Line \ref{algline:makeClique}), we perform the parallel BFS by
\cite{UllmanY91} from each of the centers.
By Lemma \ref{lem:ESTCluster}, the diameter of the input graphs to
recursive calls after the top level is bounded by $O(n^{\gamma_2}\log n)$.
Therefore the parallel BFS only need to be ran for $O(n^{\gamma_2}\log n)$
levels.
This gives a total depth of $O(n^{\gamma_2}\log n\log^*n)$
and work of $O(\rho m)$ per level.
Summing over $O(\log n)$ levels of recursion gives
$O(n^{\gamma_2}\log^2n\log^*n)$ depth
and $O(\rho m \log n) = O(\epsilon^{-\delta}m\log^{1+\delta}n)$ work.
%%
%%We now bound the work performed by our algorithm.
%%Since the recursive calls work on disjoint parts of the graph, computing the
%%exponential start time clustering takes $O(m)$ work in each level and
%%$O(m\log_\rho n)$ work overall.
%%Next observe that are at most $\rho$ big clusters in any recursive call, 
%%thus computing all-pair shortest distances among the centers of big clusters
%%does $O(\rho m)$ work in one level, and
%%$O(\rho m\log_\rho n)=O(\epsilon^{-\delta}m\log^{1+\delta}n)$ in total.
\QED

The unweighted version of Theorem~\ref{thm:mainHopSet} then follows from
Theorem \ref{thm:unweightedHopSet} by setting $\delta=1+\alpha$, and solving
$h=n^{\gamma_2}$ to balance the depth for hopset construction and the depth for
finding approximate distances using hopsets \cite{KleinS97}.
For a concrete example of setting these parameters,
$\delta=1.1$, $\epsilon=\frac{\epsilon'}{\log n}$, $\gamma_2=0.96$, and setting
$\gamma_1$ to some small constant leads to the following bound.

\begin{corollary}
  \label{cor:unweightedExample}
  For any constant $\epsilon' > 0$, there exists an algorithm for finding
  $(1+\epsilon')$-approximation to unweighted $s-t$ shortest path that runs in
  $O(n^{0.96}\log^2n\log^*n)$ depth and $O(m\log^{3.2}n )$ work.
\end{corollary}

\section{Hopsets in Weighted Graphs}

\label{sec:weighted}

In this section we show how to construct hopsets in weighted graphs with
positive edge weights.
We will assume that the ratio between the heaviest and the lightest edge
weights is $O(n^3)$.
This is due to a reduction similar to the one by Klein and
Subramanian~\cite{KleinS97}.
In that work, they partition the edges into categories with weights between
powers of $2$, and show that only considering edges from
$O(\log{n})$ consecutive categories suffices for approximate shortest path
computation.
This scheme can be modified by choosing categories with powers of $n$,
and then considering a constant number of consecutive categories suffices for
good approximations.
This result is summarized in the following lemma, we refer the readers to the
full version of this paper \cite{MPVX:arxiv} for the full proof.

\begin{lemma}
  \label{lem:collectionOfGraphs}
  Given a weighted graph $G=(V,E,w)$, we can efficiently construct a collection
  of graphs with $O(\abs{V})$ vertices and $O(\abs{E})$ edges in total,
  such that the edge weights in any one of these graphs are within $O(n^3)$ of
  each other.
  Furthermore, given a shortest path query, we can map it to a query on one
  of the graphs whose answer is a $(1-\epsilon)$-approximation for the
  original query.
\end{lemma}

Recall that the parallel BFS of \cite{UllmanY91} conducts the search level by
level, and divides the work of each level between the processors.
So a simple adaptation of parallel BFS to weighted graphs can lead to depth
linear in path lengths, which can potentially be big even though
the number of edge hops is small.
To alleviate this we borrow a rounding technique from \cite{KleinS97}.
%The main difficulty is that the weighted BFS advances its search frontier in
%increment of the minimum edge weight.
%If edges of weight $1$ and $n$ exists at the same time, the search would take
%$n$ steps to traverse an edge of weight $n$.
%This is necessary to compute exact distances, but if we are willing to pay a
%small amount of distortion, we can round the edge weights up so that the search
%advances much faster.
The main idea is to round up small edge weights and pay a small amount of
distortion, so that the search advances much faster.
%%Note that the guarantees of the weighted hop-set construction gives $O(d)$ diameter
%%as long as edge weights are integers.
%%This is because a lower bound of $1$ on edge length means a path of length $d$ can
%%be composed of at most $d$ edges.
%%As a result, the rounding step can be viewed as adjusting edge weights so that
%%small weights are simply ignored.

Suppose we are interested in a path $p$ with at most $k$ edges whose weight
is between $d$ and $c d$.
We can perturb the weight of each edge additively by $\frac{\zeta d}{k}$ without
distorting the final weight by more than $\zeta d$.
This value serves as the ``granularity'' of our rounding, which we denote using $\hat w$:
%%Suppose we are interested in a path of size at most $k$ and of weight approximately $d$.
%%We can set
\begin{align*}
  \hat w=\frac{\zeta d}{k}
\end{align*}
for some $0<\zeta<1$ and round the edge weights $w(e)$ to $\tilde w(e)$
\begin{align*}
  \tilde w(e)=\Ceil{\frac{w(e)}{\hat w}}.
\end{align*}
Notice that this rounds edge weights to multiples of $\hat w$.
The properties we need from this rounding scheme is summarized in the following lemma.

\begin{lemma}[Klein and Subramanian \cite{KleinS97}]
  \label{lem:KleinSrounding}
  Given a weighted graph and a number $d$.
  Under the above rounding scheme, any shortest path $p$ with size at most $k$ and weight $d\le
  w(p)\le cd$ for some $c$ in the original graph now has weight $\tilde
  w(p)\le\ceil{ck/\zeta}$ and $\hat w\cdot\tilde w(p)\le(1+\zeta)w(p)$.
\end{lemma}

%%To justify our assumption on having an estimate on the $st$ distance, we notice that we can easily
%%obtain an estimates that is off by a fact of at most $n$:
%%
%%\begin{lemma}
%%  \label{lem:distanceEstimate}
%%  Given a weighted graph and a pair of vertices $s,t\in V$, we can obtain an estimate
%%  $d\le\dist(s,t)\le nd$ in $O(\log n)$ depth and $O(m)$ work.
%%\end{lemma}
%%
%%\Proof
%%We compute a minimum spanning tree using Bor\r{u}vka's algorithm in parallel, then look for the
%%heaviest edge on the unique path between $s$ and $t$ on the MST and let $e'$ be that edge with
%%weight $w'$.
%%Since the $st$ path on the tree has size at most $n-1$, $nw'$ is obviously an upper bound on
%%$\dist(s,t)$. Furthermore $w'$ must also be a lower bound of $\dist(s,t)$, otherwise we can $e'$
%%with the an edge from the $st$ shortest and get a spanning of smaller weight.
%%\QED

Thus we only need to run weighted parallel BFS for $O(ck\zeta^{-1})$ levels to
recover $p$, giving a depth of $O(ck\zeta^{-1}\log n)$.
Therefore, if we set $c=n^\eta$ for some $\eta<1$, and since the edge weights are within $O(n^3)$
of each other, we can just try building hopsets using $O(3/\eta)$ estimates, incurring a factor of
$O(3/\eta)$ in the work.
%%Thus we assume that we are given a weighted graph with $n$ vertices
%%and a guess $d$ on the length of the $st$ path of interest such
%%that $d\le\dist(s,t)\le n^\eta d$.
%%We round the edge weights by letting $\alpha=\zeta d/n$ and work
%%with the new weights from now on.
As one of the values tried satisfies $d \leq w(p) \leq c d$,
Lemma~\ref{lem:KleinSrounding} gives that if $\zeta$ is set to $\epsilon / 2$,
an $(1 + \epsilon/2)$-approximation of the shortest path in the rounded
graph is in turn an $(1 + \epsilon)$-approximation to the shortest path in the
original graph.
Therefore, from this point on we will focus on finding an
$(1 + \epsilon)$-approximation of the shortest path in the rounded graph
with weights $\tilde{w}(e)$.
In particular, we have that all edge weights are positive integers,
and the shortest path between $s$ and $t$ has weight
$O(n^{1+\eta}/\zeta) = O(n^{1 + \eta} / \epsilon)$.

%%Since this rounding scheme we always overestimates the distances, so we can
%%construct in parallel a number of hop sets using different estimates
%%$w',cw',c^2w',\cdots,c^{\log_cn}w'=nw'$, and take the minimum answer.  If we
%%set $c=n^\eta$ for some $\eta<1$, then we only need to try $1/\eta$ estimates,
%%incurring a factor of $1/\eta$ in the work.  Thus we can assume that we are
%%given a weighted graph with $n$ vertices and a guess $d$ on the length of the
%%$st$ path of interest such that $d\le\dist(s,t)\le n^\eta d$.  We round the
%%edge weights by letting $\alpha=\zeta d/n$ and work with the new weights from
%%now on.  In particular, we can assume that the minimum edge weight is $1$ and
%%the distance between $s$ and $t$ under the new weights will be
%%$n^{1+\eta}/\zeta$.

\begin{theorem}
  \label{thm:weightedHopSet}
  \begin{sloppypar}
  %Given a graph $G=(V,E)$ with $\abs{V}=n$ and $\abs{E}=m$, a pair of vertices
  %$s$ and $t$.
  %, and an estimate on the $st$ distance $d\le\dist(s,t)\le n^\eta d$.
  For any constants $\delta>1$ and $\gamma_1<\gamma_2<1$, we can construct a
  $(\epsilon \log n,h,O(n))$-hopset on a graph with $n$ vertices and $m$ edges
  in expected
  $O((n/\epsilon)^{\gamma_2}\log^2n\log^*n)$ depth and
  $O(m\log^{1+\delta}n\epsilon^{-\delta})$ work, where
  $h=n^{1+1/\delta+\eta+\gamma_1(1-1/\delta)-\gamma_2}/\epsilon^{1-\gamma_2}$.
  \end{sloppypar}
\end{theorem}

\Proof
Since the edge weights are within a polynomial of each other, we can build
$O(1/\eta)$ hopsets in parallel for all values of $d$ being powers of
$n^\eta$. For any pair of vertices $s$ and $t$, one of the value tried will
satisfy $d\le\dist(s,t)\le n^\eta d$.
Given such an estimate, we first perform the rounding described above, then we
run Algorithm \ref{alg:hopSet} with $\beta=(n/\epsilon)^{-\gamma_2}$ and
$n_{\mit{final}}=n^{\gamma_1}$.
The exponential start time clustering in Line \ref{algline:ESTCluster} takes place
in the weighted setting, and Line \ref{algline:makeClique} becomes a weighted
parallel BFS.
The correctness of the hopset constructed follows from Lemma
\ref{lem:hopAndDistortion}, Lemma \ref{lem:hopSetSize}, and the fact that
$\dist(s,t)=O(n^{1+\eta}/\epsilon)$ by the rounding.
Specifically, the expected hop count is
\begin{align*}
  n^{1/\delta}n_{\mit{final}}^{1-1/\delta}\beta d
  &\le
  n^{1/\delta}n^{\gamma_1(1-1/\delta)}
  \left(\frac{n}{\epsilon}\right)^{-\gamma_2}\frac{n^{1+\eta}}{\epsilon}
  \\&=
  n^{1+1/\delta+\eta+\gamma_1(1-1/\delta)-\gamma_2}/\epsilon^{1-\gamma_2}
\end{align*}
and the expected distortion is $O(\epsilon d)$. By Markov's inequality, the
probability of both of these exceeding four times their expected values is at
most $1/2$.

The number of recursion levels is still bounded by $\log_\rho n$.
Since the $\beta$s only increase, according to Lemma \ref{lem:ESTCluster} we spend
$O((n/\epsilon)^{\gamma_2}\log n\log^*n)$ depth in each level of the recursion and
$O((n/\epsilon)^{\gamma_2}\log^2n\log^*n)$ overall in Line~\ref{algline:ESTCluster}.
Since our decomposition is laminar, we spend $O(m)$ work in each level and
$O(m\log n)$ overall in Line~\ref{algline:ESTCluster}.
Again, Line \ref{algline:makeStar} can be incorporated into the decomposition
with no extra cost.

Since the diameter of the pieces below the top level is bounded by
$\beta^{-1}\log n=(n/\epsilon)^{\gamma_2}\log n$ and the minimum edge weight is
one, Line \ref{algline:makeClique} can be implemented by weighted parallel BFS
in depth $O((n/\epsilon)^{\gamma_2}\log n\log^*n)$ in one level and
$O((n/\epsilon)^{\gamma_2}\log^2n\log^*n)$ in total.
The work done by the weighted parallel BFS is $O(\rho m)$ per level and $O(\rho
m\log n)=O(m \log^{1+\delta}n\epsilon^{-\delta})$ in total.
%By Lemma \ref{lem:ESTCluster}, Line \ref{lem:ESTCluster} takes
%$O((n/\epsilon)^\delta\log n\log_\rho n)$ depth and $O(m\log_\rho n)$ work
%in total.
%Similarly to the unweighted case, since the diameter of the pieces are less
%then $\epsilon(n/\zeta)^\delta$ in all levels below the top level, Line
%\ref{algline:makeClique} takes $O(\epsilon(n/\zeta)^\delta\log n\log_\rho n)$
%depth and $O(\rho m\log_\rho n)$ work in total.
\QED

Theorem~\ref{thm:mainHopSet} then follows from Theorem \ref{thm:weightedHopSet} 
% and Lemma~\ref{lem:usehopset} 
by adjusting the various parameters.
Again, to give a concrete example, we can set $\delta=1.1$,
$\epsilon=\epsilon'/(\log n)$, $\gamma_2=0.96$, and set $\gamma_1$ and $\zeta$
to some small constants to obtain the following corollary

\begin{corollary}
  \label{cor:weightedExample}
  For any constant error factor $\epsilon'$, there exists an algorithm for
  finding $(1+\epsilon')$-approximation to weighted $s$-$t$ shortest path that
  runs in $O(n^{0.96}\log^2n\log^*n)$ depth and $O(m\log^{3.2}n)$ work in a
  graph with polynomial edge weight ratio.
\end{corollary}

Notice that with our current scheme it is not possible to push the depth under
$\tilde O(\sqrt n)$ as the hop count becomes the bottle neck.
A modification that allows us to obtain a depth of $\tilde O(n^\alpha)$ for
arbitrary $\alpha>0$ at the expense of incurring more work can be found in the
full version of this paper \cite{MPVX:arxiv}.

\section*{Acknowledgements}

Miller and Xu are supported by the National Science Foundation under grant
number CCF-1018463 and CCF-1065106.
Peng was partially supported by a Microsoft Research Ph.D. Fellowship.
Vladu was supported by a Teaching Assistantship provided by the MIT Department
of Mathematics.
 
We thank Timothy Chu and Yiannis Koutis for helpful discussions.
 
% \begin{spacing}{0.7}
%   \begin{small}
    \bibliographystyle{alpha}
    \newcommand{\etalchar}[1]{$^{#1}$}

%   \end{small}
% \end{spacing}

\pagebreak

\begin{appendix}

\section{Deferred Proofs}
% \section{Deferred Proofs from Section~\ref{sec:ESTCluster}}
% is ref in section title broken with sig-alternate?

\label{sec:proofs}

%% For completeness, pseudocode of our decomposition scheme is given in
%% Algorithm~\ref{alg:partition}.

%% \begin{algo}[ht]
%% \textsc{Partition}$(G,\beta)$
%% \vspace{0.05cm}

%% \underline{Input:}
%% Graph $G = (V, E, w)$, parameter $\beta$.

%% \underline{Output:} Decomposition of $G$.

%% \begin{algorithmic}[1]
%% 	\STATE{For each vertex $u$, pick $\delta_u$ independently from the exponential distribution $\exp(\beta)$.}
%% 	\STATE{Assigning each vertex $v$ to the vertex $u$ that minimizes $\dist(u, v) - \delta_u$}
%% 	\STATE{For each vertex $u$ with one or more $v$s assigned to them, create a cluster including these vertices with $u$ as center.}
%% \end{algorithmic}

%% \caption{Partition Algorithm Using Exponentially Shifted Shortest Paths}
%% \label{alg:partition}

%% \end{algo}

We now show the properties of the exponential start time clustering routine
from Section~\ref{sec:ESTCluster} in more detail.

\Proofof{Lemma~\ref{lem:ESTCluster}}

The bound on diameter of the clusters follows from taking a union bound on the
maximum value of $\delta_u$ at vertices:
\begin{align*}
  \Pr\left[\delta_{\text{max}}>\frac{k\log n}\beta\right]
  &\le
  \sum_{v\in V}\Pr\left[\delta_v>\frac{k\log n}\beta\right]
  \\&=
  n\cdot\exp\left(-\beta\cdot\frac{k\log n}\beta\right)
  \\&=
  \frac{1}{n^{k-1}}.
\end{align*}

To compute the output of \textsc{ESTCluster} efficiently, we can add a
super-source and connect it to vertex $u$ via an edge of length $\delta_u$,
then we build a shortest path tree in increasing order of distance.
The clusters then correspond to the subtrees below the super-source.
We can construct this shortest path tree level by level in
$O(\frac{\log{n}}{\beta})$ steps, taking only the integer part of the
$\delta_u$s into consideration with arbitrary tie breaking in the search.
Since we assume the minimum edge weight is $1$, this modification in
implementation can be shown to have negligible effect on the probabilistic
guarantee from Lemma~\ref{lem:intersectSubset} (see Theorem~2 from \cite{ShunDB14}).
The overhead of $O(\log^{*}n)$ per search level comes from the overhead of CRCW
PRAM~\cite{GilMV91}.
We remark that this factor of $\log^{*}n$ depends on the model of parallelism,
but is standard in parallel BFS algorithms~\cite{KleinS97}.
It is $O(1)$ in the OR CRCW PRAM model, and can be bounded by $O(\log{n})$ in
most models of parallelism.
\QED

\Proofof{Lemma~\ref{lem:intersectSubset}}

Let $B$ be a subgraph of $G$ with center $c$ and radius $r$.
From $c$'s point of view, the algorithm can be seen as a race between all the
vertices to $c$: vertex $v$ starts its race at time
$\delta_{\text{max}}-\delta_v$, and arrives at $c$ at time $d(v,c)+\delta_{\text{max}}-\delta_v$.
In particular, the winner of this race will include $c$ in its cluster.
For $B$ to intersect $k$ or more clusters, the first $k$ arrivals at $c$ must
be within $2r$ units of time of each other.
Since $\delta_{\text{max}}$ is a common term in everyone's arrival time, we can
drop it and flip the sign to obtain a quantity $Y_v=\delta_v-d(v,x)$,
for each vertex $v$.
Notice that it is just an exponential random variable with some constant
offset.
The event we are interested in then becomes: the $k$ largest $Y_v$'s are within
$2r$ of each other.

We will use the law of total probability for continuous random variables.  
Let $S$ vary over subsets of $V$ of size $k-1$, $u\in V\setminus S$, and
$\alpha$ a fix real number.
Let $E_{S,u,\alpha}$ be the event that $Y_u=\alpha$ and $Y_v\ge\alpha$ if $v\in
S$ and $Y_v<\alpha$ if $v\not\in S$.
That is, set $S$ represents the first $k-1$ arrivals, and $u$ is the $k$th
arrival at time $\alpha$.
Clearly, ranging over all possible $S$, $u$ and $\alpha$ gives a paritition of
the probability space.
Thus it suffices to show
\begin{gather*}
  \Pr[\text{$Y_v\le\alpha+2r$ for all $v\in S$}\mid E_{S,u,\alpha}]
  \le
  (1-\exp(-2r\beta))^{k-1}
\end{gather*}
for any fixed $S$, $u$ and $\alpha$.

By independence of the $Y_v$s we have that 
\begin{align*}
  &\ \Pr[\text{$Y_v\le\alpha+2r$ for all $v\in S$}\mid E_{S,u,\alpha}]
  \\=&\ 
  \prod_{v \in S} \Pr[Y_v \leq \alpha +2r \mid\alpha\le Y_v].
\end{align*}
For each $v\in S$,
\begin{align*}
  &\ \Pr[Y_v\le\alpha+2r\mid\alpha\le Y_v]
  \\=&\ 
  \Pr[\delta_v\le\alpha+2r+d(v,c)\mid\alpha+d(v,c)\le X_v].
\end{align*}
There are two cases to consider.
If $\alpha+d(v,c)\le0$, then by the definition of the exponential distribution
\begin{align*}
  &\ \Pr[\delta_v\le\alpha+2r+d(v,c)\mid\alpha+d(v,c)\le X_v]
  \\\le&\ 
  \Pr[\delta_v\le2r]
  \\=&\ 
  1-\exp(1-2r\beta).
\end{align*}
If $\alpha+d(v,c)>0$, using the memoryless property of the exponential
distribution, we have
\begin{align*}
  &\ \Pr[\delta_v\le\alpha+2r+d(v,c)\mid\alpha+d(v,c)\le X_v]
  \\=&\ 
  \Pr[\delta_v\le2r]
  \\=&\ 
  1-\exp(1-2r\beta).
\end{align*}
This finishes the proof.
\QED

%% Let $x$ be an arbitrary vertex in $X$.
%% From $x$'s point of view, the algorithm can be seen as a race between all the
%% vertices to $x$: vertex $v$ starts its race at time
%% $\delta_{\text{max}}-\delta_v$, and arrives at $x$ at time
%% $d(v,x)+\delta_{\text{max}}-\delta_v$.
%% In particular, the winner of this race will include $x$ in its cluster.
%% For $X$ to be shattered by more than $s$ clusters, the first $s$ arrivals at
%% $x$ must be within $2r$ units of time of each other.
%% Since $\delta_{\text{max}}$ is a common term in everyone's arrival time, we can
%% drop it and flip the sign to obtain a quantity $Y_v=\delta_v-d(v,x)$,
%% for each vertex $v$.
%% Notice that it is just an exponential random variable with some constant
%% offset.
%% The event we are interested in then becomes: the $s$ largest $Y_v$'s are within
%% $2r$ of each other.
%% If we condition on the $s$th largest $Y_v$, by the memoryless property of the
%% exponential distribution, this happens with probability at most
%% $(1-\exp(2r \beta))^{k-1}$.
%% \QED

\section{Preprocessing to Create Instances with Polynomially Bounded Edge Weights}
\label{sec:preprocessing}

Here we describe the reduction needed for the assumption of the edge
weights being polynomially bounded in Section~\ref{sec:weighted}.
We will give a way to transform a graph $G$ into a collection of graphs where
the ratio between the maximum and minimum edge weights is at most $O(n^3)$ in
each graph.
The total size of the collection of graphs is on the order of the original graph,
and given any query, we can map it to a query in one of the graphs in the
collection efficiently.
The technique presented is similar to the scheme used by Klein and
Sairam~\cite{KleinS92}.
They partition edges into categories with weights between consecutive powers
of $2$, and show that only considering edges from $O(\log{n})$ consecutive
categories suffice for approximate shortest path computation.
We modify this scheme slightly by choosing categories by powers of $n$,
and show that picking a constant number of consecutive categories suffice.

%This allows us to view the hop set construction as a preprocessing step, and query complexity can be
%improved if we are willing to spend more work in preprocessing.

%When generalizing our hop set construction from unweighted graphs to weighted graphs, we somewhat
%weakened our result in the sense that the hop set construction in a weighted graph had to depend on
%the pair of vertices being queried.
%With unweighted graphs, the hop set construction presented in Section~\ref{sec:unweighted} is
%already a preprocessing step, since $n$ is an upper bound on any shortest path and thus we can build
%%%the hop set independent of the query.
%%With weighted graphs, the potentially huge variations between the edge weights makes it difficult to
%%balance between diameter of the hop set and the distortion.
%Given a query and with suitable rounding, we can bring the edge
%weights to the appropriate scale for that particular query,
%but this scale is not necessarily appropriate for other shortest path distances.
%Too large a scale can introduce too much distortion, while too small a scale
%does not give any speedup.
%We solve this problem by partitioning all possible queries into
%a constant number of groups.
%This will allow us to apply the algorithm developped in Section \ref{sec:weighted} to each group
%independently and preprocess the graph without needing first to know the query.

We will divide the edges into categories according to their weights
so that weights between edges in categories that are more than $2$
apart differ significantly.
If the shortest path needs to use an edge in a heavier category,
any edges in lighter weight categories can be discarded with
minor distortion.
Thus setting edge weights in these lower categories to $0$ does not change
the answer too much.
As the graph is undirected, this is equivalent to constructing the quotient
graph formed by contracting these edges.
%In our analysis, it's often useful to view the graph between
%these clusters without taking the internal structure of clusters into account.
We then show that the total size of these quotient graphs over all
categories is small.
This allows us to precompute all of them beforehand, and use hopsets for
one of them to answer each query.
To simplify notations when working with these quotient graphs,
we use $G / E'$ to denote the quotient graph formed by contracting
a subset of edges $E' \subseteq E$.

Given a weighted graph $G=(V,E,w)$, we may assume that the minimum
edge weight is $1$ by normalizing by the minimum weight.
Then we group the edges into categories as follows:
\begin{align*}
E_i = \left\{ e \in E \mid (n / \epsilon)^i \le w(e) < (n / \epsilon)^{i + 1}
\right\}.
\end{align*}
As the contractions are done to all edges belonging to some lower category,
they correspond to prefixes in this list of categories.
We will denote these using $P_j$, $P_j=\bigcup_{i=0}^j E_i$.
Also, let $q(1),\cdots,q(k)$ be the indices of the non-empty categories in $G$.
Contracting $E_1, E_2 \ldots $ in order leads to a laminar decomposition
of the graph, which we formalize as a hierarchical category decomposition:

\begin{definition}
  \label{def:laminarTree}

  A hierarchical weight decomposition is defined inductively as follows.
  \begin{itemize}
    \item The vertices form the leaves. For convenience we say that the leaves form the
      $0$\textsuperscript{th} level and define $E_{q(0)}=\emptyset$.
    \item Given the $j$\textsuperscript{th} level whose nodes represent connected components of
      $G[E_{q(j)}]$, we form the $(j+1)$\textsuperscript{th} level by adding a node for each
      connected components of $G[E_{q(j+1)}]$, and make it the parent of the components in
      $G[E_{q(j)}]$ it contains.
  \end{itemize}
\end{definition}

\begin{lemma}
  \label{lem:makeLaminarTree}
  A hierarchical weight decomposition can be computed in $O(\log^3n)$
depth and $O(m\log{n})$ work.
\end{lemma}

\Proof
We first compute the non-empty categories $E_{q(1)},\cdots,E_{q(k)}$
where $k\le m$.
Then we perform divide and conquer on the number of weight categories.
Let $E_j$ be the median weight class, the connected components
of $G[E_j]$ can be computed using the graph connectivity algorithm
by Gazit~\cite{Gazit93} in $O(\log{n})$ depth and $|E_j|$ work.
We then recurse on each connected components and also on
the quotient graph where all the components of $G[E_j]$ are collapsed to a point.
\QED

This then allows us to prove Lemma~\ref{lem:collectionOfGraphs} at the start of
Section~\ref{sec:weighted} about only working with graphs with polynomially
bounded edge weights.

\Proofof{Lemma~\ref{lem:collectionOfGraphs}}
We first construct the decomposition tree from Definition \ref{def:laminarTree}.
Once we have the tree, given a query on the distance between $s$ and $t$,
we can find their least common ancestor (LCA) in the tree using parallel
tree contraction.
Let $j$ be the level the LCA of $s$ and $t$ is in, then we claim that we only
need to consider edges in $E_{q(j-1)}\cup E_{q(j)}\cup E_{q(j+1)}$.
Since the LCA is in $j$\textsuperscript{th} level, the $s-t$ shortest path
uses at least one edge, say $e_j$, from $E_{q(j)}$.
By definition, for any edge $e_{j - 2} \in P_{q(j-2)}$,
we have $(n/\epsilon)w(e_{j - 2})\le w(e_j)$.
Since the $s-t$ path can have at most $n-1$ edges, setting lengths of edges in
$E_{q(j-2)}$ to $0$ incurs a multiplicative distortion of at most $\epsilon$.
Moreover, edges in level $j+2$ and above have weights at least
$(n/\epsilon)w(e')$, and since $s$ and $t$ is in one connected
components of $G[E_{q(j)}]$, no edge in level $j+2$ and above can be
part of the $s-t$ shortest path.

Consider the induced subgraph $G[P_{q(j+1)}]$ and its quotient
graph where all edges in $P_{q(j-1)}$ are collapsed to points:
$G[P_{q(j+1)}] / P_{q(j-1)}$.
Let $s'$ be the component in $G[E_{q(j-1)}]$ containing $s$
and let $t'$ be the component that contains $t$.
By the above argument, the shortest path between $s'$ and $t'$
in $G[P_{q(j+1)}] / P_{q(j-1)}$ is an $(1-\epsilon)$-approximation
for the $s-t$ shortest path in $G$.
Lemma~\ref{lem:makeLaminarTree} allows us to build the graphs
$G[P_{q(j+1)}]/ P_{q(j-1)}$ for all $j$ as part of the decomposition
tree construction without changing the total cost of constructing the
hopsets.
Each edge of $G$ appears at most three times in these quotient graphs,
however the number of vertices is equal to the size of the decomposition tree.
We trim down this number by observing that any chain in the tree
of length more than three can be shortened to length three by
throwing out the middle parts as they will never be used for any query.
This gives an overall bound of $O(\abs{V})$.
\QED

%%
%%
%%\Proofof{of Lemma~\ref{lem:polybounded}}
%%This gives us a way to construct hop sets on weighted graph
%%independent of the queries.
%%We construct a collection of graphs using Theorem \ref{thm:collectionOfGraphs},
%%where edge weights are within a factor of $O(n^3)$ of each other in any of
%%these graphs.
%%Then in each of these graphs, we construct in parallel constant
%%number of hop sets using Theorem~\ref{thm:weightedCorrectness}.
%%Using the bounds from Theorem~\ref{thm:collectionOfGraphs}~and ~\ref{thm:weightedCorrectness}, this can be done in $O(\log^3{n})$ depth
%%and $O(m\log{n})$ work.
%%\QED

\section{Obtaining Lower Depth}
\label{sec:lowerdepth}

We now show that the depth of our algorithms can be reduced
arbitrarily to $n^\alpha$ for any $\alpha>0$ in ways similar to the
Limited-Hopset algorithm by Cohen~\cite{Cohen00}.
So far, we have been trying to approximate paths of potentially $n$ hops with
paths of much fewer hops.
%<<<<<<< .mine
Consider the bound from Theorem~\ref{thm:unweightedHopSet},
which gives a hop count of $n^{1+1/\delta+\gamma_1-\gamma_2}$
for $\delta>1$ and $\gamma_1<\gamma_2<1$.
The the first factor of $n$ is a result of handling path containing up to
$n$ hops directly.
We now show a more gradual scheme that gradually reduces the
length of these paths.
Instead of reducing the hop-count of paths containing up to $n$
edges, we approximate $n^{2\eta}$-hop paths with ones
containing $n^{\eta}$ hops for some small $\eta$.
This routine can be applied to a longer path with $k$ hops, by breaking
it into $k n^{- 2\eta}$ ones of $n^{2 \eta}$ hops each and
apply the guarantee separately.
If the guarantee holds deterministically, we get an approximation
with $k n^{-\eta}$ hops.
Repeating this for $1 / \eta$ steps would then lead to a low depth
algorithm.
However, the probabilistic guarantees of our algorithms makes it
necessary to argue about the various piece simultaneously.
This avoids having probabilistic bounds on each piece separately,
but rather one per weight class.
%
%
%Even thought the $st$ shortest path may have size up to $O(n)$, this can be
%overcome by repeatedly building hop sets on top of others $O(1/\theta')$ times.
%%We were able to obtain a hop count of $n^{1+1/\delta+\gamma_1-\gamma_2}$ for
%%$\delta>1$ and $\gamma_1<\gamma_2<1$, where $n^{\gamma_2}$ also appears in the
%%depth of hop set construction.
%%Analyzing our rounding scheme and recalling Lemma \ref{lem:hopAndDistortion},
%%we see that the factor of $n$ in the hop count is a result of trying to handle
%%paths of size $O(n)$.
%%Now suppose that we are only interested in building hop sets that allow us to
%%approximate paths of size $n^{\theta+\theta'}$ by paths size $n^{\theta}$ for
%%some small $\theta$ and $\theta'$.
%%We show that this can be done faster.
%%Even thought the $st$ shortest path may have size up to $O(n)$, this can be
%%overcome by repeatedly building hop sets on top of others $O(1/\theta')$ times.
%%>>>>>>> .r4026
%Another point to notice that is we need to cover paths of size $n^\theta$ of
%\emph{any} weight.
%To do so, we group paths according to their weights into buckets of the form
%$[d,n^\eta d]$ for some small $\eta$.
%There are $1/\eta$ buckets and we handle them in parallel using $d$ as an
%estimates for each bucket.

\begin{lemma}
\label{lem:limitedhop}
Given a graph $G = (V, E, w)$, let $p_1 \ldots p_t$ be a collection
of disjoint paths hidden from the program such that
each $p_i$ has at most $k = n^{2 \eta}$ hops and weight between
$d$ and $d n^{\eta}$.
For any failure probability $p_{failure}$, we can construct in
$O(n^{\eta} / \epsilon)$ depth and
$O(m \log^{2 + \frac{2}{\eta}} n / \epsilon)$ work a set of at most
$O(n^{1 - \frac{\eta}{2} } )$ edges $E'$ such that with probability at least $1 - p_{failure}$
there exist paths $p'_1 \ldots p_t'$ such that:
\begin{enumerate}
	\item $p'_i$ starts and ends at the same vertices as $p_i$.
	\item The total number of hops in $p'_1 \ldots p'_t$ can be bounded by
		$t n^{\eta}$.
	\item $\sum_{i = 1}^{t} w(p'_i) \leq (1 + \epsilon) \sum_{i = 1}^{t} w(p_i)$.
\end{enumerate}
\end{lemma}

\Proof

We use the rounding scheme and construction for weighted paths from
 Section~\ref{sec:weighted}.
We first round the edge weights with $\hat{w}= \epsilon d n^{-2\eta}$.
As the paths have at most $k = n^{2\eta}$ edges, the guarantees of
Lemma~\ref{lem:KleinSrounding} gives that the lengths of paths
are distorted by a factor of $(1 + \epsilon)$.
Furthermore, this rounding leaves us with integer edge weights such
that the total length of each path is at most $d = n^{3 \eta} / \epsilon$.

We can then call Algorithm~\ref{alg:hopSet} on the rounded graph with
the following parameters:
%starting/ending conditions:
%%=======
%%Fix any shortest path $p$.
%%As in Section \ref{sec:weighted}, we assume polynomial edge weight ratio and we
%%have an estimate $d\le w(p)\le n^\eta d$. 
%%We first round the edge weights with $\alpha=\zeta d/n^{\theta+\theta'}$ to
%%bring the weight of $p$ from $w(p)$ to $\hat
%%w(p)=n^{\theta+\theta'+\eta}/\zeta$.
%%We then call Algorithm \ref{alg:hopSet} with
%%>>>>>>> .r4026
\begin{align*}
  \delta = \frac{2}{\eta},\;\;
\beta_0 =\left(\frac{n^{3\eta}}{\epsilon}\right)^{-1} =
\frac{1}{d},\;\;
n_{\mit{final}} =n^{\frac{\eta}{2}},\;\;
\epsilon' = \frac{\epsilon}{\log{n}}.
%%  \hspace{1cm}
%%  \text{and}
%%  \hspace{1cm}
%%  n_{final}=n^\gamma.
\end{align*}

%%=======
%%Along the same lines as the proof of Theorem \ref{thm:weightedHopSet} and Lemma
%%\ref{lem:hopAndDistortion}, we can show that this
%%takes $O(\epsilon(n^{\theta+\eta}/\zeta)\log n\log_Kn)$ depth and $O(m\log^\alpha n\log n)$ work.
%%Furthermore any path $p$ of size at most $n^{\theta+\theta'}$ and weight $d\le
%%w(p)\le n^\eta d$ can be approximated by new path of size at most
%%$n^{1/\alpha+\gamma+\eta}$.
%%>>>>>>> .r4026

By an argument similar to the proof of Theorem~\ref{thm:weightedHopSet}
and Lemma~\ref{lem:hopAndDistortion}, this takes
$O((n^{2\eta}/\epsilon)\log n\log_Kn \epsilon)$ depth and
$O(m\log^{1 + \delta} n / \epsilon') = O(m\log^{2 + \frac{2}{\eta}} n /
\epsilon)$ work.
Furthermore, for each $p_i$, the expected number of pieces that it is
partitioned into is:
\begin{align*}
  n^{\frac{1}{\delta}} \beta_0 d n_{\mit{final}}
 = n^{\frac{\eta}{2}} n^{\frac{\eta}{2}}
 = n^{\eta}
\end{align*}
and if we take all shortcuts through centers of big clusters,
the expected distortion is:
\begin{align*}
O(\log_{\rho_{n}} n \epsilon' d)
& = \epsilon d.
\end{align*}
Applying linearity of expectation over all $t$ paths gives
that the expected total number of hops is $t n^{\eta}$,
and the expected additive distortion is $\epsilon d t$.
As both of these values are non-negative, Markov's inequality
the probability of any of these exceeding $\frac{2}{p_{failure}}$
of their expected value is at most $p_{failure}$.
Therefore, adjusting the constants and $\epsilon$ accordingly
gives the guarantee.
Finally, the number of edges in the hopset can be bounded
by $n^{1 - \eta} \log^{\frac{4}{\eta}} n \leq n^{1 - \frac{\eta }{2}}$.
\QED

Then it suffices to run this routine for all values of $d$
equaling to powers of $n^{\eta}$.
The fact that edge weights are polynomially bounded
means that this only leads to a constant factor overhead in work.
Running this routine another $n^{\eta}$ times gives the hop-set
paths with arbitrary number of hops.

\begin{theorem}
  \label{thm:lowDepthHopSet}
  Given a graph $G=(V,E, w)$ with polynomially bounded edge weights
  and any constant $\alpha > 0$,
  we can construct a $(\epsilon, n^{\alpha} , O(n))$-hopset for $G$ in
  $O(n^{\alpha} \epsilon^{-1})$ depth and
  $O(m\log^{O(\frac{1}{\alpha})}n \epsilon^{-1})$ work
\end{theorem}

\Proof
Let $\eta=\alpha/2$.
We will repeat the following $\frac{1}{\eta}$ times:
run the algorithm given in Lemma~\ref{lem:limitedhop} repeatedly
for all values of $d$ being powers of $n^{\eta}$, and add the
edges of the hopset to the current graph.
As the edge weights are polynomially bounded, there are $O(\frac{1}{\eta^2})$
invocations, and we can choose the constants to that they can all succeed
with probability at least $1/2$.
In this case, we will prove the guarantees of the final set of edges
by induction on the number of iterations.

Consider a path $p$ with $k$ hops.
If $k \leq n^{2 \eta}$, then the path itself serves as a $k$-hop equivalent.
Otherwise, partition the path into pieces with $n^{2 \eta}$ hops,
with the exception of possibly the last $n^{2 \eta}$ edges.
Consider these subpaths classified by their weights.
The guarantees of Lemma~\ref{lem:limitedhop} gives that
each weight class can be approximated with a set of paths
containing $n^{- \eta}$ as many edges.
Putting these shortcuts together gives that there is a path $p'$
with $k n^{-\eta}$ hops such that $w(p') \leq (1 + \epsilon) w(p)$.
Since $p'$ has fewer edges, applying the inductive
hypothesis gives that $p'$ has an equivalent in the final graph
with $n^{2\eta}$ hops that incurs a distortion of
$(1 + (\log_{n^{\eta}}{k}  - 1) \epsilon)$.
Multiplying together these two distortion factors
gives that this path also approximates $p$ with distortion
$(1 + \log_{n^{\eta}}{k} \epsilon)$.
As $k \leq n$ and $\eta$ is a constant, replacing $\epsilon$
with $\eta \epsilon$ gives the result.
\QED

%%
%%ALso from Lemma \ref{lem:hopAndDistortion}, we see that one round introduces
%%$(1+\epsilon\log_Kn)$ relative distortion.
%%Since we repeately build $1/\theta'$ hop sets, we introduced
%%$O\left((1+\epsilon)^{1/\theta'}\right)$ relative distortion in total.
%%
%%As an concrete example, we can set $\theta=1/2$, $\alpha=2$ and $\theta'$, $\gamma$ and $\eta$ to
%%some small constants to get about $\tilde O(\sqrt{n})$ depth and $O(m\log^3n)$ work.

\end{appendix}

\end{document}